\documentclass[12pt]{iopart} 
\usepackage{iopams}
\usepackage{graphicx}
\usepackage{caption}
\usepackage{color}
\usepackage{cite}
\usepackage{bm}
\expandafter\let\csname equation*\endcsname\relax
\expandafter\let\csname endequation*\endcsname\relax
\usepackage{amsmath,amssymb,amsfonts}
\usepackage{multirow}%
\usepackage{amsthm}%
\usepackage{mathrsfs}%
\usepackage[title]{appendix}%
\usepackage{xcolor}%
\usepackage{textcomp}%
\usepackage{manyfoot}%
\usepackage{booktabs}%
\usepackage{algorithm}%
\usepackage{algorithmicx}%
\usepackage{algpseudocode}%
\usepackage{listings}
\usepackage[normalem]{ulem}
\usepackage{cancel}
\begin{document}
 
\title[Learning rate matrix and information-thermodynamic trade-off relation]{Learning rate matrix and information-thermodynamic trade-off relation}

\author{Kenshin Matsumoto, Shin-ichi Sasa and Andreas Dechant}
\address{Department of Physics, Kyoto University, Kyoto 606-8502, Japan}

\eads{\mailto{matsumoto.kenshin.73z@st.kyoto-u.ac.jp}, \mailto{sasa@scphys.kyoto-u.ac.jp} and \mailto{dechant.andreas.3z@kyoto-u.ac.jp}}

\begin{abstract}
  Non-equilibrium systems exchange information in addition to energy.
  In information thermodynamics, the information flow is characterized by the learning rate, which is not invariant under coordinate transformations. To formalize the property of the learning rate under variable transformations, we introduce a learning rate matrix. This matrix has the learning rates as its diagonal elements and characterizes the changes in the learning rates under linear coordinate transformations. The maximal eigenvalue of the symmetric part of the learning rate matrix gives the maximal information flow under orthogonal transformations. Furthermore, we derive a new trade-off relation between the learning rate and the heat dissipation of a subsystem. 
  Finally, we illustrate the results using analytically solvable yet experimentally feasible models.
\vspace*{1ex}\\
\textbf{Keywords:} stochastic thermodynamics 
\end{abstract}

\text{} Revised version of July 24, 2025
\section{Introduction}\noindent
Physical systems exchange information in addition to energy. The study of Maxwell's demon has evolved into the field of information thermodynamics 
 \cite{Sagawa,parrondo2015thermodynamics}, integrating information theory \cite{coverthomas} and stochastic thermodynamics \cite{Sekimoto2010}. By measuring and applying feedback control to a system, Maxwell's demon uses information to extract work from an equilibrium heat bath or to cool the system. Furthermore, information is of crucial importance to biological organisms, which process information as a means of reacting to changes in their environment. In the work of information thermodynamics, fundamental limits have been formulated for the robustness of signal transduction against external fluctuations and the thermodynamic efficiency of information flow \cite{cellar_information,Sensory_capacity,ito2015maxwell}. In recent studies, universal relations between the fluctuations of information flows and the entropy production rate have been derived by incorporating information flows into the thermodynamic uncertainty relation \cite{TUR,Tanogami_Saito_2023}.\\
One way of quantifying the information flows between two physical systems is the so-called learning rate. In the second law of information thermodynamics, accounting for the information flow in terms of the learning rate ensures that the entropy production rate is positive for each system \cite{langevin_information,conti_information,Multipartite_information_flow,penocchio2022information}. When the evolution of a system with degrees of freedom $\boldsymbol{x} = \lbrace x_i \rbrace^{i=1}_{N}$ follows the Fokker-Planck equation (see below for details), the learning rate is defined by decomposing the time-derivative of the mutual information into contributions from each subsystem $x_i$ \cite{non-bipartite}:
\begin{equation}
     l^{i}_t
     =
     \int d \boldsymbol{x} J_{t,i}(\boldsymbol{x}) \frac{\partial}{\partial x_i} \ln \left(\frac{p_t(\boldsymbol{x})}{p^i_t(x_i)}\right) \quad (i=1,2,\dots,N),
\end{equation}
where $p_t(\boldsymbol{x})$ is the probability density, $p^i_t(x_i)$ is the marginal density of $x_i$, and $J_{t,i}$ is the probability current of $X_i$. From this definition, the learning rate depends on the choice of coordinates; in particular, an orthogonal transformation of the coordinates results in a different learning rate. This raises the question of whether there exist geometric invariants that characterize the information flow in the system. \\
In this work, we introduce a learning rate matrix in order to characterize the behavior of the learning rate with respect to variable transformations. For a given choice of coordinates, the learning rates appear as the diagonal elements of this matrix. The behavior of the learning rate matrix under orthogonal transformations allows us to define a ``learning capacity'', which represents the upper limit of the information flow in the system. This quantity is given as the maximal eigenvalue of the symmetric component of the learning rate matrix. Furthermore, a matrix inequality derived from the learning rate matrix allows us to formulate a new trade-off relation between the learning rate and the heat dissipation of a subsystem. This inequality obtained in this study gives a nonzero lower bound on the sum of the dissipation and the learning rate, and is tighter than the second law of information thermodynamics.\\
The remainder of this paper is organized as follows. In Section 2, we introduce some fundamental information-theoretic quantities and briefly review the second law of information thermodynamics for Langevin systems. In addition, we introduce the learning rate matrix as an extension of the learning rate. In Section 3, we describe two special choices of the coordinates and the maximal value of the learning rate, constituting the first main result of this paper.
In Section 4, we derive a trade-off relation for the learning rate, which is the second main result of this paper. This inequality is tighter than the second law of thermodynamics and implies that, while a nonzero learning rate is necessary to obtain apparently negative dissipation for a subsystem, very large information flows generally lead to more dissipation. In Section 5, we demonstrate our results using two specific examples: a Brownian gyrator and a model of two interacting particles.\\

\section{Setup and definition of learning rate matrix}\noindent
\subsection{Overdamped Langevin equation}
We consider $N$ coupled overdamped Langevin equations for the dynamical variables $\boldsymbol{x}(t)=(x_1(t),\ldots,x_N(t))$:
\begin{equation}
\label{Langevin_equation}
    \dot{x}_i(t) = F_{t,i}(\boldsymbol{x}(t))+\sum_j B_{t,ij}(\boldsymbol{x}(t))\cdot \xi_j(t),
\end{equation}
where $i=1,\cdot \cdot \cdot,N$, $F_{t,i}(\boldsymbol{x})$ is a drift vector. Let $M$ be the number of noise components. $\boldsymbol{B}_t$ is the matrix of size $N\times M$, and it may depend on the system variables. $\boldsymbol{\xi}(t)$ denotes mutually independent Gaussian white noise $\langle \xi_i(t) \xi_j(s) \rangle=\delta_{ij}\delta(t-s)$, and $\{\cdot\}$ represents the Ito product \cite{gardiner1985handbook}. These dynamics can be described by the Fokker-Planck equation for the probability density $p_t(\boldsymbol{x})$ and current $\boldsymbol{J}_t(\boldsymbol{x})$:
\begin{equation}
\label{Fokker_Planck}
    \partial_t p_t(\boldsymbol{x})
    =
    -
    \sum_{i=1}^{N} \partial_{x_i} J_{t,i}(\boldsymbol{x})
    =
    -
    \sum_{i=1}^{N} \partial_{x_i} (\nu_{t,i}(\boldsymbol{x})p_t(\boldsymbol{x})),
\end{equation}
where the probability current $J_{t,i}(\boldsymbol{x})$ is given by
\begin{equation}
    J_{t,i}(\boldsymbol{x})
    =
    \left(
    F_i(\boldsymbol{x}) - \frac{1}{2}\sum_j \partial_{x_j} D_{ij}(\boldsymbol{x})
    \right)p_t(\boldsymbol{x}).
\end{equation}
Additionally, the local mean velocity is defined as $\boldsymbol{\nu}_t(\boldsymbol{x})\equiv\boldsymbol{J}_t(\boldsymbol{x})/p_t(\boldsymbol{x})$. The diffusion matrix $\boldsymbol{D}_t\equiv \boldsymbol{B}_t\boldsymbol{B}_t^{\mathrm{T}}$ is positive definite.
\\
\subsection{Entropy production rate, mutual information, and learning rates}
The entropy of the system is given by the Shannon entropy \cite{Shannon,coverthomas}:
\begin{equation}
    S_t
    =
    -\int d \boldsymbol{x} p_t(\boldsymbol{x}) \ln p_t(\boldsymbol{x}).
\end{equation}
The rate of change of the system entropy is identified as the time-derivative of the Shannon entropy:
\begin{equation}
    \dot{S}_t
    =
    -\int d \boldsymbol{x} \frac{\partial}{\partial t}p_t(\boldsymbol{x}) \ln p_t(\boldsymbol{x}),
\end{equation}
which vanishes in the steady state, where $\partial_t p_t(\boldsymbol{x}) = 0$. The total entropy change in the surrounding environment is defined as
\begin{equation}
    \dot{S}_t^{\mathrm{env}}
    =
    \langle
    (\boldsymbol{F}_t-\boldsymbol{\nabla}\boldsymbol{D}_t)^{\mathrm{T}}\boldsymbol{D}_t^{-1}\boldsymbol{\nu}_t
    \rangle,
\end{equation}
which is directly related to the heat dissipation, and is referred to as the dissipation here and in the following. Therefore, the total entropy production rate is given as
\begin{equation}
\begin{split}
\label{total_entorpy_production_rate}
    \sigma_t
    &=
    \dot{S}_t+\dot{S}_t^{\mathrm{env}}\\
    &=
    \langle \boldsymbol{\nu}_t^{\mathrm{T}} \boldsymbol{D}_t^{-1} \boldsymbol{\nu}_t
    \rangle,
\end{split}
\end{equation}
which has a non-negative value and is zero only if the detailed balance condition $J_{i,t}=0, {}^\forall i\in \{1,\ldots ,N\}$ is satisfied. This non-negativity corresponds to the second law of thermodynamics.
Let us now derive the second law for subsystems.
Just as for the total system, the entropy of the subsystem is given by the Shannon entropy of the marginal probability density:
\begin{equation}
    S_t^i
    =
    -\int d \boldsymbol{x}_i p^i_t(x_i) \ln p^i_t(x_i),
\end{equation}
where $p^i_t(x_i)$ is the marginal probability density of $x_i$. The entropy change in the surrounding environment caused by the heat flow from subsystem $i$ is defined as 
\begin{equation}
\begin{split}
    \dot{S}^{i,\mathrm{env}}_t
    &=
    \langle
    (F_{t,i}-\nabla_i D_{t,ii})
    (D_{t,ii})^{-1}
    \circ
    \dot{x}_i
    \rangle,
\end{split}
\end{equation}
where $\{\circ\}$ represents the Stratonovitch product. Similarly to the total system, $\dot{S}^{i,\mathrm{env}}_t$ is regarded as the dissipation in the subsystem. Here, we assume the diffusion matrix to be diagonal, which corresponds to the multipartite condition, because the noise of the subsystems is uncorrelated \cite{Multipartite_information_flow}. 
Following \cite{SUNHAN198026}, we introduce the mutual information for multiple variables as
\begin{equation}
    I_t
    =
    \int d \boldsymbol{x}
    p_t(\boldsymbol{x})
    \ln \left(
    \frac{p_t(\boldsymbol{x})}{p^1_t(x_1)p^2_t(x_2) \cdots p^N_t(x_N)}
    \right).
\end{equation}
As a Kullback-Leibler divergence, the mutual information is non-negative. The mutual information is equal to zero if each subsystem is independent of all other subsystems, $p_t(\boldsymbol{x})=\prod_i p^i_t(x_i)$.\\
We can write the mutual information as the difference between the Shannon entropies of the marginal densities and the Shannon entropy of the joint density:
\begin{equation}
    I_t
    =
    \sum_{i=1}^N S^i_t-S_t .
\end{equation}
The time-derivative of the mutual information can be calculated using the Fokker-Planck equation (\ref{Fokker_Planck}) and partial integrals:
\begin{equation}
\begin{split}
    d_t I_t
    &=
    - \int d \boldsymbol{x}\boldsymbol{\nabla}^{\mathrm{T}}
    (\boldsymbol{\nu}_t(\boldsymbol{x})p_t(\boldsymbol{x}))
    \ln \left(
    \frac{p_t(\boldsymbol{x})}{p^1_t(x_1)p^2_t(x_2)\cdot \cdot \cdot p^N_t(\boldsymbol{x_N})}
    \right)\\
    &=
    \sum_{i=1}^N
    \left\langle
    \nu_{t,i} \partial_{x_i} \ln \left(  \frac{p_t}{p^i_t} \right)
    \right\rangle,
\end{split}
\end{equation}
where we have omitted the variable dependencies from the first to the second line for simplicity, and the dependencies can be understood from the explicit expressions of $\boldsymbol{\nu}$, $p_t$ and $p_t^i$.
The information flow to one variable can be quantified as the learning rate, which is defined as the decomposition of the time-derivative of the mutual information \cite{conti_information}:
\begin{equation}
    l^i_t
    =
    \left\langle
    \nu_{t,i} \partial_{x_i} \ln \left(  \frac{p_t}{p^i_t} \right)
    \right\rangle
    =
    \langle
    \nu_{t,i} \partial_{x_i} \ln p^{|i}_t
    \rangle.
\end{equation}
Here, we denote the degrees of freedom except $x_i$ by $\boldsymbol{x}_{/i}=(x_1,\ldots,x_{i-1},x_{i+1},\ldots,x_N)$ and the probability density conditioned on $x_i$ by $p^{|i}_t(\boldsymbol{x}_{/i }|x_i)=p_t/p^i_t$. The learning rate measures the rate at which the mutual information increases because of probability flows in direction $i$. The relation between the learning rate and the rate of change of the subsystem entropy is written as
\begin{equation}
    l^i_t
    =
    \left\langle
    \nu_{t,i} \partial_{x_i} \ln p_t
    \right\rangle
    +
    \dot{S}^i_t.
\end{equation}
The total entropy production rate is decomposed into the partial entropy production rate of the subsystems as $\sigma_t=\sum_i \sigma_t^i$ when the system has a multipartite structure \cite{Multipartite_information_flow}. Furthermore, the partial entropy production rate for a subsystem can be divided into the system entropy change, the entropy change caused by the heat dissipation, and the learning rate \cite{conti_information}: 
\begin{equation}
\begin{split}
\label{partial_entorpy_production_rate}
    \sigma_t^i
    &=
    \dot{S}^i_t+\dot{S}_t^{i,\mathrm{env}}-l^i_t\\
    &=
    \langle  \nu^i_t (D_{t,ii})^{-1} \nu_t^i
    \rangle.
\end{split}
\end{equation}
The partial entropy production rate satisfies $\sigma^i_t\geq 0$, which is called the second law for the subsystem. This implies that the learning rate can reduce the entropy of the subsystem and its environment.

\subsection{Introduction of learning rate matrix}
As an extension of the learning rate, we introduce the learning rate matrix as
\begin{equation}
\label{learning rate matrix}
    (\boldsymbol{L}_t)_{ij}
    \equiv
    \langle
    \nu_{t,i}\partial_j \ln p_t
    \rangle
    +
    \dot{S}^i_t \delta_{ij}.
\end{equation}
The diagonal elements of this matrix are given by the learning rates $(\boldsymbol{L}_t)_{ii}=l^i_t$. From the definition of the learning rate, the time-derivative of the mutual information corresponds to the trace of the learning rate matrix:
\begin{equation}
\label{learning_trace}
    d_t I_t
    =
    \mathrm{tr}(\boldsymbol{L}_t).
\end{equation}
In particular, in the steady state, we have
\begin{equation}
    \mathrm{tr}(\boldsymbol{L}_t)
    =
    0.
\end{equation}
The definition of the learning rate matrix satisfying (\ref{learning_trace}) would not be unique (see \ref{appendix_another_definition}).  In this paper, we use (\ref{learning rate matrix}) to clarify the relation with the time-derivative of the Shannon entropy. Differences in definitions do not affect our main results because we focus on the steady state.\\
Although the learning rates depend on the coordinates, their sum is always zero in the steady state. Therefore, the trace of the learning rate matrix is invariant under coordinate transformations.\\

\section{Learning rate matrix and learning capacity}\noindent
The purpose of this paper is to explore invariants related to information flows. In the previous section, we observed that the trace of the learning rate matrix is one such invariant. In this section, we derive other invariants by studying the behavior of the learning rate matrix under orthogonal transformations.
As the first result, we describe the behavior of information flows under orthogonal coordinate transformations. This allows two special coordinates to be identified, and we define the ``learning capacity'' as the invariant characterizing the total information flow in the system.
\subsection{Behavior of learning rates under coordinate rotations}
The learning rates depend on the choice of variables. We can formulate the dependence of the learning rates on the coordinates using the learning rate matrix. We consider the orthogonal transformation
\begin{equation}
\label{orthogonal_transformation}
    \Tilde{\boldsymbol{x}}
    =
    \boldsymbol{R}\boldsymbol{x},
\end{equation}
where $\boldsymbol{R}$ is an orthogonal transformation matrix satisfying $\boldsymbol{R}^{\mathrm{T}}\boldsymbol{R}=\boldsymbol{\mathrm{I}}$. Using the transformation of the Fokker-Planck equation, we can obtain the learning rate matrix of the new coordinate $\tilde{\boldsymbol{x}}$ as
\begin{equation}
\label{transformation_learning_matrix}
    \Tilde{\boldsymbol{L}}
    =
    \boldsymbol{R}\boldsymbol{L}\boldsymbol{R}^{-1}
\end{equation}
(see \ref{appendix_transformation}). This relation only holds in the steady state. For time-dependent systems, we also need to transform the Shannon entropy.
We discuss the two special coordinates in the following subsections.
\\
\subsection{Vanishing learning rates}
One of the two special coordinate systems in the steady state has a coordinate with no information flow. This means that the diagonal elements of the learning rate matrix are zero. In particular, for a linear Langevin equation, we can find an orthogonal transformation that results in no learing. When the system is described by the linear Langevin equation, the drift term is assumed to be 
\begin{equation}
    \boldsymbol{F}_t(\boldsymbol{x})
    =
    -\boldsymbol{K}_t \boldsymbol{x} + \boldsymbol{k}_t,
\end{equation}
and the diffusion matrix $\boldsymbol{D}_t(\boldsymbol{x})=\boldsymbol{D}_t$. As shown in \ref{appendix_linear_case}, if the initial distribution is Gaussian, the learning rate matrix is given by
\begin{equation}
\label{learning_rate_matrix_linear_system}
    \boldsymbol{L}_t
    =
    \frac{1}{2}
    (\boldsymbol{A}_t-\dot{\boldsymbol{\Xi}}_t) \boldsymbol{\Xi}_t^{-1} + \dot{\boldsymbol{S}}_t,
\end{equation}
where we define the skew-symmetric matrix
\begin{equation}
\label{irreversible_circulation_def}
    \boldsymbol{A}_t
    =
    (\boldsymbol{K}_t \boldsymbol{\Xi}_t - \boldsymbol{\Xi}_t \boldsymbol{K}_t^{\mathrm{T}}),
\end{equation}
called irreversible circulation \cite{irreversible_circulation}, $\boldsymbol{\Xi}_t$ is the covariance matrix, and $\boldsymbol{S}_t$ is a diagonal matrix with the Shannon entropies as its diagonal elements $S_{t,ij}=S_t^i\delta_{ij}$. In the steady state, we can further write (\ref{learning_rate_matrix_linear_system}) as
\begin{equation}
    \boldsymbol{L}_{\mathrm{st}}
    =
    \frac{1}{2}
    \boldsymbol{A}_{\mathrm{st}}
    \boldsymbol{\Xi}^{-1}_{\mathrm{st}}.
\end{equation}
Because $\boldsymbol{\Xi}_{\mathrm{st}}^{-1}$is a positive definite matrix, it can be diagonalized by an orthogonal matrix
\begin{equation}
    \boldsymbol{G}
    \boldsymbol{\Xi}_{\mathrm{st}}
    \boldsymbol{G}^{\mathrm{T}}
    =
    \mathrm{diag}(\lambda^{\Xi}_k) \quad \text{with}  \quad
    \boldsymbol{G}^{\mathrm{T}}\boldsymbol{G}
    =
    \boldsymbol{\mathrm{I}},
\end{equation}
where $\lambda^{\Xi}_k>0$ are the eigenvalues of $\boldsymbol{\Xi}_{\mathrm{st}}$. Applying this to the above, we have
\begin{equation}
    \boldsymbol{G}
    \boldsymbol{L}_{\mathrm{st}}
    \boldsymbol{G}^{\mathrm{T}}
    =
    \frac{1}{2}
    \boldsymbol{G}
    \boldsymbol{A}_{\mathrm{st}}
    \boldsymbol{G}^{\mathrm{T}}
    (
    \boldsymbol{G}
    \boldsymbol{\Xi}_{\mathrm{st}}
    \boldsymbol{G}^{\mathrm{T}}
    )^{-1}.
\end{equation}
Because the orthogonal transformation of a skew-symmetric matrix is again skew-symmetric, the right-hand side is the product of a skew-symmetric matrix and a diagonal matrix; in particular, all diagonal elements vanish. Thus, the same orthogonal transformation diagonalizes $\boldsymbol{\Xi}_{\mathrm{st}}$ and causes all diagonal elements of $\boldsymbol{L}_{\mathrm{st}}$ (i.e., the learning rates) to vanish. Therefore, in linear systems, diagonalizing the covariance matrix eliminates correlations, implying that no information flow exists between uncorrelated systems.
\\
\subsection{Maximal learning rate: learning capacity}
The other of the two special coordinate systems has a coordinate with the maximal learning rate. The maximal learning rate and the coordinate under orthogonal transformations (\ref{orthogonal_transformation}) can be obtained not only for linear Langevin systems but also for general Langevin systems. We can separate the learning rate matrix into symmetric and skew-symmetric parts:
\begin{equation}
    \boldsymbol{L}_t
    =
    \boldsymbol{L}_t^{\mathrm{sym}}
    +
    \boldsymbol{L}_t^{\mathrm{anti}}.
\end{equation}
Here, the symmetric part $\boldsymbol{L}_t^{\mathrm{sym}}$ and the skew-symmetric part $\boldsymbol{L}_t^{\mathrm{anti}}$ are defined as
\begin{equation}
    \boldsymbol{L}_t^{\mathrm{sym}}
    \equiv
    \frac{1}{2}(\boldsymbol{L}_t+\boldsymbol{L}_t^{\mathrm{T}}),\quad
    \boldsymbol{L}_t^{\mathrm{anti}}
    \equiv
    \frac{1}{2}(\boldsymbol{L}_t-\boldsymbol{L}_t^{\mathrm{T}}).
\end{equation}
Because any orthogonal transformation of a skew-symmetric matrix is again skew-symmetric, only the symmetric part of the learning rate matrix contributes to the diagonal elements. Therefore, the largest diagonal element of the symmetric part gives the largest diagonal element of the learning rate matrix. Furthermore, as a consequence of the Schur-Horn theorem \cite{Horn_Johnson_1985}, the diagonal elements of the symmetric part cannot be larger than the maximal eigenvalue: 
\begin{equation}
\begin{split}
    &\sum_{i=1}^{n}
    d_i 
    \leqq
    \sum_{i=1}^{n}
    \lambda_i\quad \mathrm{for}\quad n=1,\cdot\cdot\cdot,N-1,\\
    &\sum_{i=1}^{N}
    d_i 
    =
    \sum_{i=1}^{N}
    \lambda_i,
\end{split}
\end{equation}
where $\{d_i\}_{i=1}^N$ and $\{\lambda_i\}_{i=1}^N$ are sets of diagonal elements
and eigenvalues arranged in non-increasing order. Thus, the orthogonal transformation that diagonalizes the symmetric part simultaneously gives the largest and smallest possible diagonal elements (see \ref{smallest_eigenvalue}).
Here, we define the learning capacity $l_c$ as the largest eigenvalue of the symmetric part of the learning rate matrix. This represents the maximum capacity of the learning rate under orthogonal transformations in the steady state. The maximal possible learning rate is achieved in the direction of the eigenvector corresponding to the largest eigenvalue. This gives the variable that has the largest information flow. For non-orthogonal transformations, we cannot easily obtain the learning capacity and its coordinate because the symmetric and skew-symmetric parts are mixed. The symmetric part of the transformed learning rate matrix is not equal to the transformed symmetric part of the original matrix, and thus the eigenvalues change under the transformation. Furthermore, when non-orthogonal transformations are included, the diagonal elements of the learning rate matrix can become arbitrarily large or small.

\section{Learning rates and dissipation}\noindent
As the second main result of this study, we derive the trade-off relation for the learning rates using matrix inequalities.
\\
\subsection{Lower bound on entropy production rate from learning rate matrix}
The total entropy production rate is given by (\ref{total_entorpy_production_rate}), which can be rewritten as
\begin{equation}
\label{total_entropy_production_local}
    \sigma_t
    =
    \langle(\boldsymbol{\nu}_t-\langle\boldsymbol{\nu}_t\rangle)^{\mathrm{T}}
    \boldsymbol{D}^{-1}_t
    (\boldsymbol{\nu}_t-\langle\boldsymbol{\nu}_t\rangle)
    \rangle
    +
    \langle\boldsymbol{\nu}_t\rangle^{\mathrm{T}}
    \langle\boldsymbol{D}^{-1}_t \rangle
    \langle\boldsymbol{\nu}_t\rangle
\end{equation}
using $\langle  \langle\boldsymbol{\nu}_t\rangle^{\mathrm{T}} \boldsymbol{D}^{-1}_t (\boldsymbol{\nu}_t-\langle\boldsymbol{\nu}_t\rangle)   \rangle=0$. Because the first term is a convex function of the entries of $\boldsymbol{D}_t(\boldsymbol{x})$, Jensen's inequality implies that
\begin{equation}
\label{total_entropy_production_Jensen_bound}
    \sigma_t
    \geq
    \mathrm{tr}
    \left(
    \langle \boldsymbol{D}_t \rangle^{-1}
    \boldsymbol{\Xi}_{\nu,t}
    \right)
    +
    \langle\boldsymbol{\nu}_t\rangle^{\mathrm{T}}
    \langle\boldsymbol{D}^{-1}_t \rangle
    \langle\boldsymbol{\nu}_t\rangle,
\end{equation}
where $\boldsymbol{\Xi}_{\nu,t}$ is the covariance matrix of the local mean velocity, defined as
\begin{equation}
    (\boldsymbol{\Xi}_{\nu,t})_{ij}
    =
    \langle
    (\nu_{t,i}-\langle\nu_{t,i}\rangle)
    (\nu_{t,j}-\langle\nu_{t,j}\rangle)
    \rangle.
\end{equation}
Next, to derive the lower bound, we introduce the Fisher information matrix as
\begin{equation}
\label{Fisher_information}
    (\boldsymbol{\mathcal{I}}_t)_{ij}
    =
    \langle \partial_{x_i} \ln p_t \partial_{x_j} \ln p_t \rangle.
\end{equation}
Both $\boldsymbol{\Xi}_{\nu,t}$ and $\boldsymbol{\mathcal{I}}_t$ are symmetric and positive semi-definite matrices. The relation between these matrices and the learning rate matrix is given by
\begin{equation}
\begin{split}
\label{learning_matrix_Fisher_local_velocity}
    \boldsymbol{\Xi}_{\nu,t}
    \geq
    (\boldsymbol{L}_t-\dot{\boldsymbol{S}}_t)
    \boldsymbol{\mathcal{I}}_t^{-1}
    (\boldsymbol{L}_t-\dot{\boldsymbol{S}}_t)^\mathrm{T}
\end{split}
\end{equation}
(see \ref{appendix_learning_matrix_Fisher_local_velocity}). Combining (\ref{total_entropy_production_Jensen_bound}) and (\ref{learning_matrix_Fisher_local_velocity}), we have
\begin{equation}
\label{total_entropy_production_lower_bound}
    \sigma_t
    \geq
    \mathrm{tr}
    \left(
    \langle \boldsymbol{D}_t \rangle^{-1}
    (\boldsymbol{L}_t-\dot{\boldsymbol{S}}_t)
    \boldsymbol{\mathcal{I}}_t^{-1}
    (\boldsymbol{L}_t-\dot{\boldsymbol{S}}_t)^{\mathrm{T}}
    \right)
    +
    \langle\boldsymbol{\nu}_t\rangle^{\mathrm{T}}
    \langle \boldsymbol{D}^{-1}_t \rangle
    \langle\boldsymbol{\nu}_t\rangle.
\end{equation}
Therefore, the learning rate matrix together with the  Fisher information provides a lower bound on the entropy production rate.
The Fisher information quantifies the spatial variation of the probability density; thus, sharp spatial features of the probability density result in a large Fisher information. By the Cramer-Rao inequality, a small variance in position implies a large Fisher information. Therefore, the above inequality implies  that, under constant entropy production rate, a highly accurate estimation is accompanied by a large information flow. Furthermore, when the system is described by linear Langevin equations, the lower bound becomes an equality (see \ref{appendix_linear_case}).
\subsection{Off-diagonal elements contribute to the lower bound on entropy production rate}
We now explain the importance of the off-diagonal elements of the learning rate matrix. The first term on the right-hand side of (\ref{total_entropy_production_lower_bound}) is a convex function of the entries of $\boldsymbol{L}_t-\dot{\boldsymbol{S}}_t$, as shown by
\begin{equation}
\begin{split}
    &\partial^2_{(\boldsymbol{L}_t-\dot{\boldsymbol{S}}_t)_{ij}}
    \mathrm{tr}
    \left(
    \langle \boldsymbol{D}_t \rangle^{-1}
    (\boldsymbol{L}_t-\dot{\boldsymbol{S}}_t)
    \boldsymbol{\mathcal{I}}_t^{-1}
    (\boldsymbol{L}_t-\dot{\boldsymbol{S}}_t)^{\mathrm{T}}
    \right)\\
    &=
    (\langle\boldsymbol{D}_t\rangle^{-1})_{ii}
    (\boldsymbol{\mathcal{I}}^{-1})_{jj}
    >
    0.
\end{split}
\end{equation}
Consequently, the lower bound on the entropy production rate is a strictly increasing function of the magnitude of the entries of $\boldsymbol{L}_t-\dot{\boldsymbol{S}}_t$. Therefore, any nonzero entry of $\boldsymbol{L}_t-\dot{\boldsymbol{S}}_t$ increases the lower bound on the entropy production rate. This implies that, although the change in mutual information depends only on the diagonal elements of $\boldsymbol{L}_t$ through (\ref{learning_trace}), the off-diagonal elements also carry information about the entropy production rate in the system.\\
\subsection{Trade-off relation for learning rates}
A similar calculation can be performed for a system consisting of two subsystems. We now split the degrees of freedom $\boldsymbol{x}$ into two disjoint sets, $\boldsymbol{x}=(\boldsymbol{y},\boldsymbol{z})$. We assume that the diffusion matrix is block-diagonal:
\begin{equation}
    \boldsymbol{D}_t
    =
    \left(
    \begin{matrix}
        \boldsymbol{D}^y_t(\boldsymbol{y})&0\\
        0&\boldsymbol{D}^z_t(\boldsymbol{z})
    \end{matrix}
    \right).
\end{equation}
This assumption is equivalent to a bipartite structure with respect to $X$ and $Y$.
In this case, we can formally split the entropy production rate into two positive parts as
\begin{equation}
\label{separation_etropy_production_rate}
    \sigma_t
    =
    \sigma^y_t + \sigma^z_t
    =
    \langle (\boldsymbol{\nu}^y_t)^{\mathrm{T}} (\boldsymbol{D}_t^y)^{-1} \boldsymbol{\nu}_t^y
    \rangle
    +
    \langle (\boldsymbol{\nu}^z_t)^{\mathrm{T}} (\boldsymbol{D}_t^z)^{-1}  \boldsymbol{\nu}_t^z
    \rangle.
\end{equation}
We write the learning rate matrix as 
\begin{equation}
    \boldsymbol{L}_t
    =
    \left(
    \begin{matrix}
        \boldsymbol{L}_t^{yy}&\boldsymbol{L}_t^{yz}\\
        \boldsymbol{L}_t^{zy}&\boldsymbol{L}_t^{zz}
    \end{matrix}
    \right),
\end{equation}
and the Fisher information matrix as
\begin{equation}
    \boldsymbol{\mathcal{I}}_t
    =
    \left(
    \begin{matrix}
        \boldsymbol{\mathcal{I}}_t^{yy}&\boldsymbol{\mathcal{I}}_t^{yz}\\
        \boldsymbol{\mathcal{I}}_t^{zy}&\boldsymbol{\mathcal{I}}_t^{zz}
    \end{matrix}
    \right).
\end{equation}
The same argument leading to (\ref{total_entropy_production_lower_bound}) can be applied separately to the two terms in (\ref{separation_etropy_production_rate}). Thus, we have
\begin{equation}
\begin{split}
\label{partial_entropy_production_lower_bound}
    \sigma_t^y
    &\geq
    \mathrm{tr}
    \left(
    \langle \boldsymbol{D}_t^y \rangle^{-1}
    (\boldsymbol{L}_t^{yy}-\dot{\boldsymbol{S}}_t^y)
    (\boldsymbol{\mathcal{I}}_t^{yy})^{-1}
    (\boldsymbol{L}_t^{yy}-\dot{\boldsymbol{S}}_t^y)^{\mathrm{T}}
    \right)\\
    &+
    \langle\boldsymbol{\nu}^y_t\rangle^{\mathrm{T}}
    \langle\boldsymbol{D}_t^y\rangle^{-1}
    \langle\boldsymbol{\nu}_t^y\rangle.
\end{split}
\end{equation}
Here, $\dot{\boldsymbol{S}}^y_t$ denotes the diagonal matrix with entries $\dot{S}^k_t$ for $k \in  \boldsymbol{y}$, which is different from the Shannon entropy of the subsystem, $\mathrm{tr}(\dot{\boldsymbol{S}}^y_t) \neq \dot{S}^y_t$. Instead, the latter is given by
\begin{equation}
    \dot{S}^y_t
    =
    -
    \int d \boldsymbol{y} \ln p^y_t(\boldsymbol{y}) \partial_t p^y_t(\boldsymbol{y})
    \quad \text{with} \quad
    p^y_t(\boldsymbol{y})
    =
    \int d \boldsymbol{z} p_t(\boldsymbol{y},\boldsymbol{z}).
\end{equation}
The two quantities are related by
\begin{equation}
\dot{I}_t^y = \text{tr}(\dot{\boldsymbol{S}}^y_t) - \dot{S}^y_t,
\end{equation}
where $\dot{I}_t^y$ denotes the time-derivative of the mutual information of subsystem $\boldsymbol{y}$ and its individual degrees of freedom:
\begin{equation}
    I^y_t
    =
    \int d \boldsymbol{y}
    p^y_t(\boldsymbol{y})
    \ln \left(
    \frac{p^y_t(\boldsymbol{y})}{\prod_{k \in \boldsymbol{y}} p^k_t(x_k) }
    \right).
\end{equation}
Next, we use the Cauchy-Schwarz inequality for the trace: 
\begin{equation}
    \mathrm{tr}(\boldsymbol{A}^{\mathrm{T}}\boldsymbol{A}) 
    \geq 
    \frac{\mathrm{tr}(\boldsymbol{C}^{\mathrm{T}}\boldsymbol{A})^2}{\mathrm{tr}(\boldsymbol{C}^{\mathrm{T}}\boldsymbol{C})}.
\end{equation}
Applying this to the above expression with $\boldsymbol{A}=\sqrt{(\boldsymbol{\mathcal{I}}^{yy}_t)^{-1}}(\boldsymbol{L}_t^{yy}-\dot{\boldsymbol{S}}_t^y)^{\mathrm{T}}\sqrt{(\boldsymbol{D}^{y}_t)^{-1}}$ and $\boldsymbol{C}=\sqrt{(\boldsymbol{D}^{y}_t)}\sqrt{(\boldsymbol{\mathcal{I}}^{yy}_t)}$, we further obtain
\begin{equation}
\label{partial_entropy_production_lower_bound+Cauchy}
    \sigma_t^y
    \geq
    \frac{\mathrm{tr}
    \left(
    \boldsymbol{L}_t^{yy}-\dot{\boldsymbol{S}}_t^y
    \right)^2
    }
    {
    \mathrm{tr}
    \left(
    \langle \boldsymbol{D}_t^y \rangle \boldsymbol{\mathcal{I}}_t^{yy}
    \right)}
    +
    \langle\boldsymbol{\nu}^y_t\rangle^{\mathrm{T}}
    \langle
     \boldsymbol{D}_t^y\rangle^{-1}
    \langle\boldsymbol{\nu}_t^y\rangle.
\end{equation}
We define $\dot{S}^{y,\mathrm{env}}_t$ as the entropy change in the surrounding environment caused by the heat flow from subsystem $\boldsymbol{y}$:
\begin{equation}
\label{definition_y_env_entropy}
\begin{split}
    \dot{S}^{y,\mathrm{env}}_t
    &=
    \langle
    (\boldsymbol{F}^y_t-\boldsymbol{\nabla}_y \boldsymbol{D}^y_t)^\mathrm{T}
    (\boldsymbol{D}^y_t)^{-1}
    \circ
    \dot{\boldsymbol{y}}
    \rangle.
\end{split}
\end{equation}
The inequaly (\ref{partial_entropy_production_lower_bound+Cauchy}) expresses a non-negative lower bound on the partial entropy production rate regardless of the sign of $\dot{\boldsymbol{S}}_t^y$, while the bound from the second law (\ref{partial_entorpy_production_rate}) is zero. This is because $\langle\boldsymbol{D}^{y}_{t}\rangle$ is positive definite, and the right-hand side is greater than or equal to zero. Therefore, this inequality is tighter than the second law (\ref{partial_entorpy_production_rate}).
 Then, in the steady state, we obtain
\begin{equation}
\label{partial_entropy_production_lower_bound+Cauchy_steady_state}
    \dot{S}^{y,\mathrm{env}}_{\mathrm{st}}
    \geq
    l^y_{\mathrm{st}}
    +
    \frac{
    \left(
    l^y_{\mathrm{st}}
    \right)^2
    }
    {
    \mathrm{tr}
    \left(
    \langle \boldsymbol{D}_{\mathrm{st}}^y \rangle \boldsymbol{\mathcal{I}}_{\mathrm{st}}^{yy}
    \right)}
    +
    \langle\boldsymbol{\nu}^y_{\mathrm{st}}\rangle^{\mathrm{T}}
    \langle
    \boldsymbol{D}_{\mathrm{st}}^y\rangle^{-1}
    \langle\boldsymbol{\nu}_{\mathrm{st}}^y\rangle.
\end{equation}
This is because the rate of change of the Shannon entropy is zero in the steady state and the trace of the learning rate matrix corresponds to the learning rate:
\begin{equation}
    \mathrm{tr}
    (\boldsymbol{L}^{yy}_{\mathrm{st}})
    =
    \sum_{i \in \boldsymbol{y}}
    l^y_{\mathrm{st}}
    =
    l^y_{\mathrm{st}}.
\end{equation}
The inequality (\ref{partial_entropy_production_lower_bound+Cauchy_steady_state}) turns into an equality when the system is described by the linear Langevin equation. This inequality provides a tighter bound on the dissipation than the second law, which gives a lower bound solely with respect to the learning rate in the steady state.
Althogh a negative learning rate may allow a subsystem to absorb heat from the environment under the second law for the subsystem, this bound implies that a large negative learning rate has the opposite effect, and instead causes the subsystem to dissipate.
In general, both the learning rate and the Fisher information vary when the parameters of the system change. However, the above implies that the amount of information flow that can be effectively used to induce a negative dissipation is determined by the Fisher information, and thus the properties of the steady state probability density. Minimizing the right-hand side of \eqref{partial_entropy_production_lower_bound+Cauchy_steady_state} with respect to the learning rate yields the  lower bound
\begin{equation}
\label{lower_bound}
 \dot{S}^{y,\mathrm{env}}_{\mathrm{st}} \geq - \frac{1}{4} \mathrm{tr}
    \left(
    \langle \boldsymbol{D}_{\mathrm{st}}^y \rangle \boldsymbol{\mathcal{I}}_{\mathrm{st}}^{yy}
    \right).
\end{equation}
Thus, the maximal possible negative dissipation from a subsystem is determined by the Fisher information of its steady state probability density, whereas the learning rate is proportional to the magnitude of the local mean velocity. This inequality, which is independent of the flow contributing to dissipation, has not been reported before. For details on other bounds, see \ref{appendix_tighter_bound}.
\\

\section{Examples}\noindent
In this section, we use two specific models to demonstrate the main results presented in Sections 3 and 4.\\
We consider a two-dimensional linear Langevin equation ($i =1,2$)
\begin{equation}
\label{temperature_matrix}
    \dot{x}_i
    =
    -\sum_j K_{t,ij} x_j + k_{t,i}
    +
    \sqrt{2 T_i}\xi_i.
\end{equation}
Here, the existence of a stable steady state requires the force matrix $\boldsymbol{K}_t$ to have eigenvalues with strictly positive real parts.  For the linear system, we obtain an explicit expression for the learning rate matrix (see \ref{appendix_linear_case}).
In this section, to provide a physical interpretation of the force matrix $\boldsymbol{K}$, we write 
\begin{equation}
\label{drift_matrix}
\boldsymbol{K}
    =
    \left(
    \begin{matrix}
        k_1+\kappa&-\kappa-\delta\\
        -\kappa+\delta&k_2+\kappa
    \end{matrix}
    \right),
\end{equation}
where we assume that all parameters except for $\delta$ are positive. This means that the system has a conservative
force $\boldsymbol{f}_{\mathrm{c}}=-\partial_{\boldsymbol{x}}U(\boldsymbol{x})$, where $U(\boldsymbol{x})$ is a harmonic potential given by
\begin{equation}
\label{potential}
    U(\boldsymbol{x})
    =
    \frac{1}{2}k_1 x_1^2+\frac{1}{2}k_2 x_2^2+\frac{1}{2}\kappa(x_1-x_2)^2,
\end{equation}
and a non-conservative force $\boldsymbol{f}_{\mathrm{nc}}=\delta(x_2,-x_1)$. Here, $k_1$ and $k_2$ denote the magnitude of the trapping potential in the respective dimension, $\kappa$ denotes the strength of reciprocal coupling, and $\delta$ denotes the strength of non-reciprocal coupling. When $\delta$ is nonzero, the system has a non-equilibrium steady state. The diffusion matrix is written as
\begin{equation}
    \boldsymbol{D}
    =
    \left(
    \begin{matrix}
        T_1&0\\
        0&T_2
    \end{matrix}
    \right).
\end{equation}
For this model, we obtain the explicit form of the learning rate matrix as
\begin{equation}
    \boldsymbol{L}_{\mathrm{st}}
    =
    \frac{((\kappa-\delta)T_1-(\kappa+\delta)T_2)}{(k_1+k_2+2\kappa)\det(\boldsymbol{\Xi})}
    \left(
    \begin{matrix}
        -\Xi_{12}&
        \Xi_{11}\\
        -\Xi_{22}&
        \Xi_{12}
    \end{matrix}
    \right)
\end{equation}
with
\begin{equation}
    \begin{split}
        &\Xi_{12}=\frac{(k_1+\kappa)(\kappa+\delta)T_2+(k_2+\kappa)(\kappa-\delta)T_1}{(k_1+k_2+2\kappa)\left((k_1+\kappa)(k_2+\kappa)-(\kappa^2-\delta^2)\right)},\\
        &
        \Xi_{11}
        =
        \frac{(k_2+\kappa)(k_1+k_2+2\kappa)T_1+(\kappa+\delta)((\kappa+\delta)T_2+(\kappa-\delta)T_1)}{(k_1+k_2+2\kappa)((k_1+\kappa)(k_2+\kappa)-(\kappa^2-\delta^2))},\\
        &
        \Xi_{22}
        =
        \frac{(k_1+\kappa)(k_1+k_2+2\kappa)T_2+(\kappa-\delta)((\kappa-\delta)T_1+(\kappa+\delta)T_2)}{(k_1+k_2+2\kappa)((k_1+\kappa)(k_2+\kappa)-(\kappa^2-\delta^2))}.
    \end{split}
\end{equation}
The learning rate of $X_1$ is $(\boldsymbol{L}_{\mathrm{st}})_{11}$ and the learning rate of $X_2$ is $(\boldsymbol{L}_{\mathrm{st}})_{22}$. When 
\begin{equation}
\label{demon_conditions}
0<
    \kappa+\delta
    <
     \frac{T_1}{T_2}(\kappa-\delta),
\end{equation}
 the subsystems have a positive correlation $\Xi_{12}>0$, and $(\boldsymbol{L}_{\mathrm{st}})_{22}$ is positive because the covariance matrix is positive semi-definite ($\det(\boldsymbol{\Xi})\geq0$) and $\mathrm{tr}(\boldsymbol{K})=k_1+k_2+2\kappa>0$ for the stability condition. $X_2$ moves in a direction that increases the correlation with $X_1$, implying that $X_2$ is ``learning'' about $X_1$, effectively acting as a demon. The dissipation (\ref{definition_y_env_entropy}) from subsystem $X_1$ is 
\begin{equation}
\label{linear_partial_entropy_production}
     \dot{S}^{1,\mathrm{env}}_{\mathrm{st}}
     =
     \frac{\kappa+\delta}{r(k_1+k_2+2\kappa)}\left\{(\kappa+\delta)-(\kappa-\delta)r\right\},
\end{equation}
and this becomes negative under condition (\ref{demon_conditions}), where we define the ratio of the temperatures as $r=T_1/T_2$. This indicates that subsystem $X_1$ is cooled in the steady state. Inequality (\ref{partial_entropy_production_lower_bound+Cauchy_steady_state}) becomes an equality because the system is linear. We now specify the parameters and demonstrate the results for two examples.

\subsection{Brownian gyrator}
\begin{figure}[htbp]
\begin{minipage}{0.5\hsize}
   \centering
    \includegraphics[width=60mm]{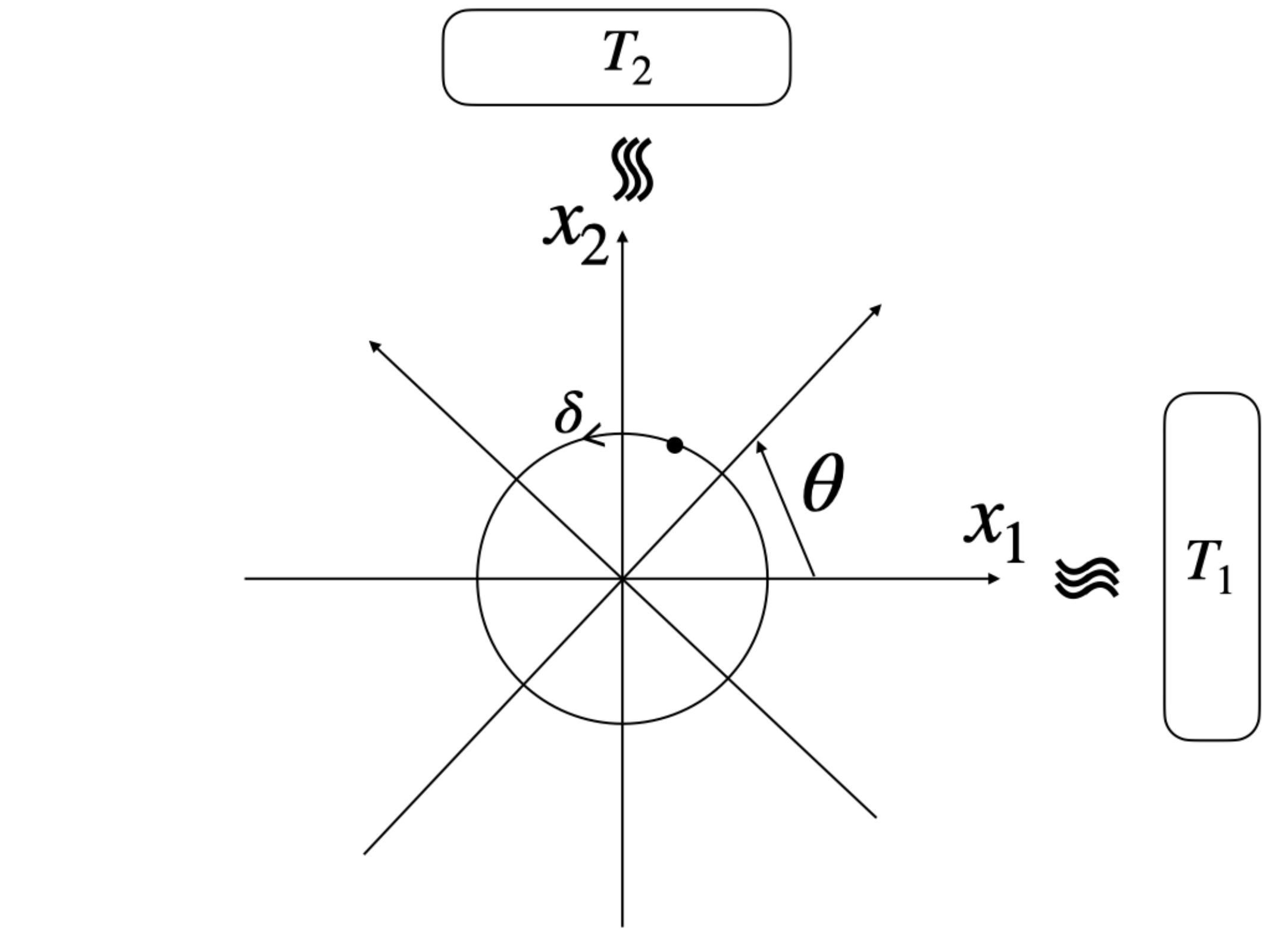}
    \captionsetup{width=0.9\linewidth,font=small}
    \caption{Sketch of a Brownian gyrator. A particle rotates under a non-conservative force in contact with heat baths $T_1$ and $T_2$. }
    \label{fig:Brownian_gyrator}
 \end{minipage}
 \begin{minipage}{0.5\hsize}
  \centering
    \includegraphics[width=60mm]{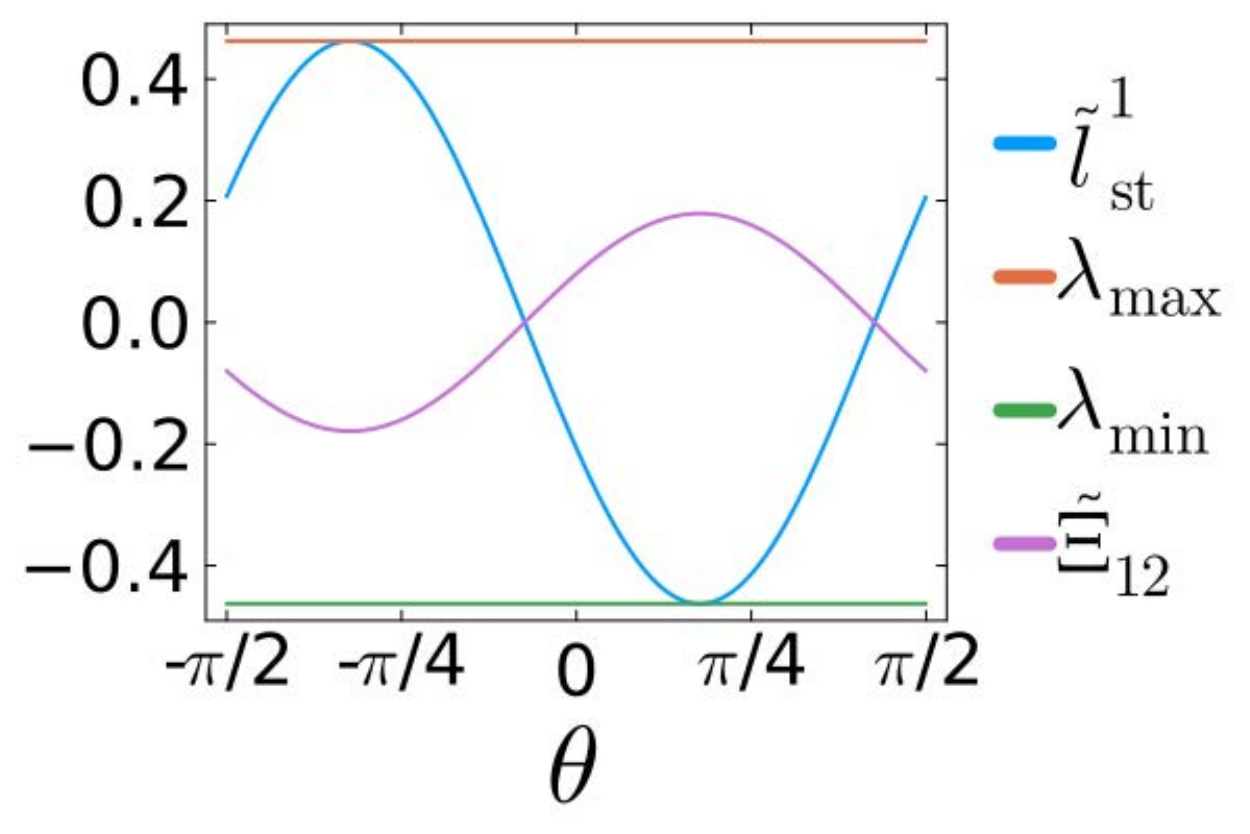}
    \captionsetup{width=0.9\linewidth,font=small}
    \caption{Orthogonal transformation of the learning rate matrix in the steady state with $T_1=0.5$, $T_2=0.1$, $k=1$, and $\delta=0.5$. $\tilde{l}^1_{\mathrm{st}}$ and $\tilde{\Xi}_{12}$ are the learning rate of $\tilde{x}_1$ and the covariance after the rotation transformation by angle $\theta$ in the steady state. $\lambda_{\mathrm{max}}$ and $\lambda_{\mathrm{min}}$ are the maximal and minimal eigenvalues.}
    \label{fig:maximal_eigenvalue}
\end{minipage}
\end{figure}

One of the analytically solvable examples of the two-dimensional case is a Brownian gyrator, which is a Brownian particle trapped in a harmonic potential and driven by a non-conservative force. This model has previously been shown to work as a heat engine \cite{Brownian_gyrator,Information_Arbitrage}. We consider a two-dimensional overdamped Brownian particle with position $\boldsymbol{x}=(x_1,x_2)$ (Figure \ref{fig:Brownian_gyrator}). The particle moves under an isotropic harmonic potential (\ref{potential}) with $\kappa=0$. The forces are given by $\boldsymbol{f}_{\mathrm{c}}=-\partial_{\boldsymbol{x}}U(\boldsymbol{x})$ and $\boldsymbol{f}_{\mathrm{nc}}=\delta(x_2,-x_1)$.
The two degrees of freedom are in contact with the baths at temperatures $T_1$ and $T_2$, respectively. For this model, the learning rate matrix is expressed as 
\begin{equation}
    \boldsymbol{L}_{\mathrm{st}}
    =
    \frac{-\delta(T_1+T_2)}{(k_1+k_2+2\kappa)\det(\boldsymbol{\Xi})}
    \left(
    \begin{matrix}
        -\Xi_{12}&
        \Xi_{11}\\
        -\Xi_{22}&
        \Xi_{12}
    \end{matrix}
    \right)
\end{equation}
with
\begin{equation}
    \begin{split}
        &\Xi_{12}=\frac{\delta(k_1T_2-k_2T_1)}{(k_1+k_2)\left(k_1k_2+\delta^2\right)},\\
        &
        \Xi_{11}
        =
        \frac{k_2(k_1+k_2)T_1+\delta^2(T_2-T_1)}{(k_1+k_2)\left(k_1k_2+\delta^2\right)},\\
        &
        \Xi_{22}
        =
        \frac{k_1(k_1+k_2)T_2+\delta^2(T_2-T_1)}{(k_1+k_2)\left(k_1k_2+\delta^2\right)}.
    \end{split}
\end{equation}
Let us rotate the measurement coordinates by the time-independent angle $\theta$. The Brownian gyrator is a heat engine that performs work through torque using the temperature difference between heat baths. In this study, we measure a subsystem, instead of the torque, while using the same setup. By applying a rotational transformation, which is orthogonal, we investigate how the information flow between variables changes. The transformation matrix for the rotation is
\begin{equation}
    \boldsymbol{R}(\theta)
    =
    \left(
    \begin{matrix}
        \cos \theta & -\sin \theta\\
        \sin \theta & \cos \theta
    \end{matrix}
    \right).
\end{equation}
The learning rate matrix in the new coordinates is given by
\begin{equation}
    \tilde{\boldsymbol{L}}
    =
    \boldsymbol{R}
    \boldsymbol{L}
    \boldsymbol{R}^{-1}
\end{equation}
according to (\ref{transformation_learning_matrix}).
Figure \ref{fig:maximal_eigenvalue} shows the change in the learning rate matrix and the covariance matrix under the rotational transformation, where we define the covariance matrix of the new coordinates as $\tilde{\boldsymbol{\Xi}}=\boldsymbol{R}\boldsymbol{\Xi}\boldsymbol{R}^{\mathrm{T}}$. As shown in the figure, the learning rate is zero when the covariance is zero and the maximum learning rate corresponds to the maximal eigenvalue. In this way, selecting coordinates that maximize the eigenvalue of the symmetric part of the learning rate matrix helps identify those where dissipation is minimized or work output is maximized. When the sign of $\delta$ is reversed, the signs of the covariance and learning rate also reverse. Because the largest and smallest eigenvalues have the same absolute value but opposite signs, their magnitudes remain unchanged under the sign reversal of $\delta$.

\subsection{Two interacting particles}
\begin{figure}[htbp]
\begin{minipage}{0.5\hsize}
  \centering
   \includegraphics[width=60mm]{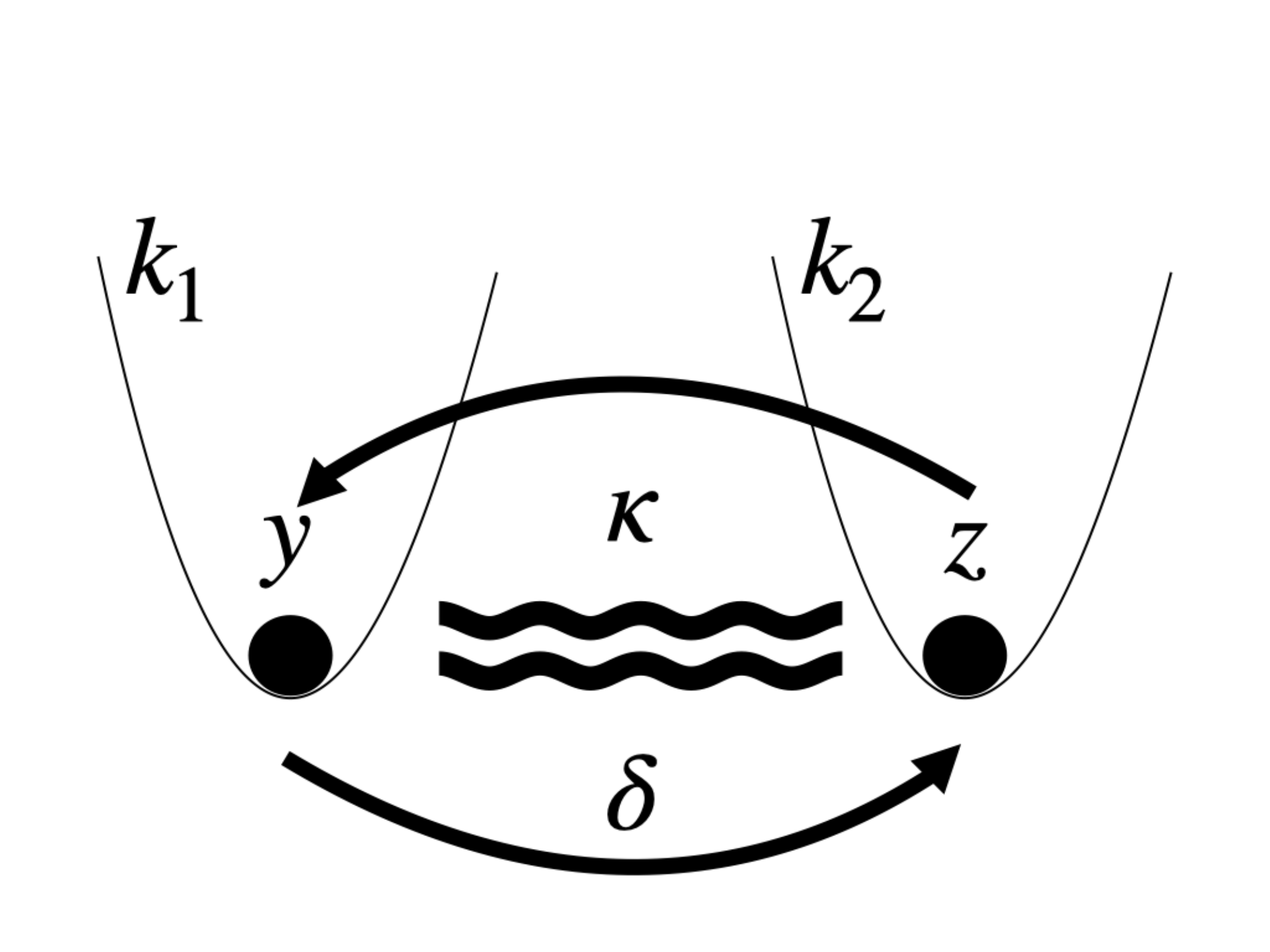}
   \captionsetup{width=0.9\linewidth,font=small}
   \caption{Sketch of two interacting particles trapped in harmonic potentials. The interaction is governed by conservative and non-conservative forces of strengths $\kappa$ and $\delta$, respectively.}
   \label{fig:two-interacting-particles}
 \end{minipage}
 \begin{minipage}{0.5\hsize}
  \centering
    \includegraphics[width=60mm]{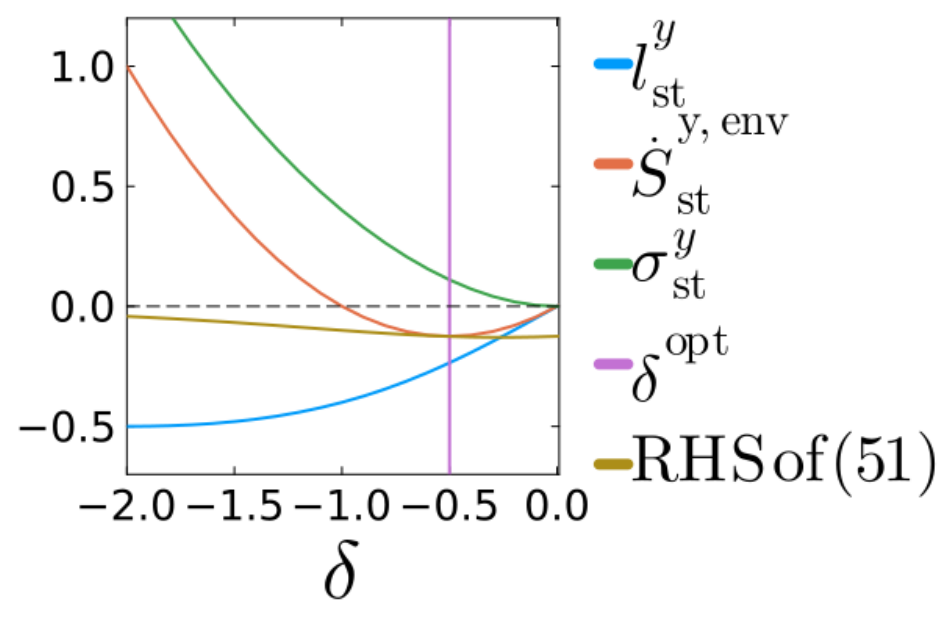}
    \captionsetup{width=0.9\linewidth,font=small}
    \caption{Learning rate and dissipation from the non-conservative force in the steady state with $k_1=k_2=T_1=T_2=\kappa=1$. $l^y_{\mathrm{st}}$ is the learning rate, $\dot{S}^{y,\mathrm{env}}_{\mathrm{st}}$ is the dissipation, and $\sigma^{y}_{\mathrm{st}}$ is the entropy production rate of subsystem $y$ in the steady state. $\delta^{\mathrm{opt}}$ is the optimal value of the strength of the non-conservative force.}
    \label{fig:over-learning}
\end{minipage}
\end{figure}

Another analytically tractable example is a model of two interacting particles. Specifically, these are two Brownian interacting particles trapped in harmonic potentials in one-dimension, and are driven by a non-conservative force (Figure \ref{fig:two-interacting-particles}). This model has been realized experimentally in optically levitated particles \cite{two_interaction}.
The particle positions are represented as $\boldsymbol{x}=(y,z)$. Similar to the Brownian gyrator, the conservative and non-conservative forces are $\boldsymbol{f}_{\mathrm{c}}=-\partial_{\boldsymbol{x}}U(\boldsymbol{x})$ and $\boldsymbol{f}_{\mathrm{nc}}=\delta(z,-y)$, where $\delta$ describes the strength of the non-conservative force. Furthermore, we set $r= 1(T_1=T_2= T)$ for simplicity.
For this model, the explicit form of the learning rate matrix is
\begin{equation}
    \boldsymbol{L}_{\mathrm{st}}
    =
    \frac{-\delta T}{(k+\kappa)\det(\boldsymbol{\Xi})}
    \left(
    \begin{matrix}
        -\Xi_{12}&
        \Xi_{11}\\
        -\Xi_{22}&
        \Xi_{12}
    \end{matrix}
    \right)
\end{equation}
with
\begin{equation}
    \begin{split}
        &\Xi_{12}=\frac{(k_1+\kappa)(\kappa+\delta)T+(k_2+\kappa)(\kappa-\delta)T}{(k_1+k_2+2\kappa)\left((k_1+\kappa)(k_2+\kappa)-(\kappa^2-\delta^2)\right)},\\
        &
        \Xi_{11}
        =
        \frac{(k_2+\kappa)(k_1+k_2+2\kappa)T+2\kappa(\kappa+\delta)T}{(k_1+k_2+2\kappa)((k_1+\kappa)(k_2+\kappa)-(\kappa^2-\delta^2))},\\
        &
        \Xi_{22}
        =
        \frac{(k_1+\kappa)(k_1+k_2+2\kappa)T+2\kappa(\kappa-\delta)T}{(k_1+k_2+2\kappa)((k_1+\kappa)(k_2+\kappa)-(\kappa^2-\delta^2))}.
    \end{split}
\end{equation}
Under this condition, the sign of the dissipation depends on the relation between the reciprocal coupling $\kappa$ and the non-reciprocal coupling $\delta$, which means this relation determines whether $Z$ works as a demon or not. Figure \ref{fig:over-learning} shows the dissipation and information flow of subsystem $Y$ and the lower bound of (\ref{lower_bound}).
Inequality (\ref{lower_bound}) is satisfied in the entire parameter region.
In the regime where $-\kappa<\delta<0$, the interaction force exceeds the non-equilibrium force, and subsystem $Z$ works as the demon. This implies that the dissipation (\ref{linear_partial_entropy_production}) from $Y$ to the environment is negative because of the information flow from $Y$ to $Z$ ($l^y_{\mathrm{st}} < 0$). The optimal value of $\delta$ is obtained by minimizing the right-hand side of (\ref{partial_entropy_production_lower_bound+Cauchy_steady_state}) with respect to the strength of the nonconservative force $\delta$. In this case, when the strength of the non-conservative force is 
\begin{equation}
    \delta^{\mathrm{opt}}
    =
    -
    \frac{\kappa}{2},
\end{equation}
the dissipation attains its smallest negative value. Therefore, in this regime, we can cool subsystem $Y$ and the right-hand side of (\ref{partial_entropy_production_lower_bound+Cauchy_steady_state}) gives the optimal strength of the non-conservative force.\\
In the regime where $\delta<-\kappa$, the non-equilibrium force surpasses the interaction strength. The magnitude of the information flow $l^y_{\mathrm{st}}$ increases further, but subsystem $Y$ has positive dissipation. This means that, instead of cooling subsystem $Y$, the demon heats the system. \\
In this way, the results obtained in this study indicate that it is possible to estimate the dissipation from statistical quantities or determine the parameters to cool a subsystem. Note that, although the inequality (\ref{partial_entropy_production_lower_bound+Cauchy_steady_state}) may provide a method for identifying parameters that minimize the dissipation, it does not guarantee that the dissipation is actually minimized at the same parameter values in general cases.

\section{Conclusion}\noindent
In this study, we addressed the issue that the learning rate, which quantifies the information flow, depends on the choice of coordinates and is not a geometric invariant that characterizes the overall information flow of the system. To explore geometric invariants, we introduced the learning rate matrix. The first invariant was found to be the trace of the learning rate matrix, which always vanishes in the steady state. The second invariant we identified was the maximal eigenvalue of the symmetric components of the learning rate matrix, which corresponds to the maximal information flow under orthogonal transformations. Furthermore, in linear systems, coordinates with no information flow can be obtained as the coordinates that diagonalize the covariance matrix, i.e., the principal component analysis coordinates. These results were demonstrated through analysis of systems where such validation is possible.
As our second result, using matrix-form inequalities, we derived a relation between the learning rate and dissipation. This inequality is tighter than the second law and implies that excessive information flow leads to dissipation.\\
In this paper, we have focused on overdamped Langevin systems. However, the proposed framework could also be applied to underdamped Langevin systems. In contrast, extending the theory to discrete systems described by master equations is not straightforward, and situations requiring variable transformations are rare in such cases. Additionally, the result that allows the extraction of coordinates with no information flow is limited to linear systems. Investigating the physical meaning of this property in nonlinear systems is left as future work.
\section*{Acknowledgments}
This work was supported by JST SPRING (Grant Number JPMJSP2110) and JSPS KAKENHI (Grant Number  JP23K22415). AD is supported by JSPS KAKENHI (Grant Number 22K13974, 24H00833 and 25K00926) and JST ERATO Grant Number JPMJER2302.

\appendix

\section{Another definition of the learning rate matrix}
\label{appendix_another_definition}
As another definition of the learning rate matrix, one may define
\begin{equation}
\label{learning rate matrix_another}
    (\boldsymbol{L}_t)_{ij}
    \equiv
    \left\langle
    \nu_{t,i}\partial_j \ln \left( \frac{p_t(\boldsymbol{x})}{p^j_t(x_j)} \right)
    \right\rangle, 
\end{equation}
which yields the same diagonal elements as (\ref{learning rate matrix}). In this paper, we focus on the steady state, where the two definitions given in (\ref{learning rate matrix}) and (\ref{learning rate matrix_another}) are equivalent. Definition (\ref{learning rate matrix_another}) differs from (\ref{learning rate matrix}) in the off-diagonal elements.  The advantage of definition (\ref{learning rate matrix_another}) is that, like the mutual information, it vanishes identically when $p_t(\boldsymbol{x})=\prod_i p^i_t(x_i)$. However, we use definition (\ref{learning rate matrix}) so that the time-derivative of the Shannon entropy appears explicitly in the inequalities. 

\section{Derivation of (\ref{transformation_learning_matrix})}
\label{appendix_transformation}
Let us consider the steady state of the Fokker-Planck equation with $\boldsymbol{x}=(x_1,\dots,x_N)\in \mathbb{R}^N$.\\
We introduce an invertible variable
transformation to $\boldsymbol{y}\in \mathbb{R}^N$:
\begin{equation}
    \boldsymbol{y}
    =
    \boldsymbol{g}(\boldsymbol{x})
    \Leftrightarrow
    \boldsymbol{x}
    =
    \boldsymbol{g}^{-1}(\boldsymbol{y}),
\end{equation}
and define the Jacobian of the transformation as
\begin{equation}
    (\boldsymbol{J}(\boldsymbol{x}))_{ij}
    =
    \partial_{x_j}g_i (\boldsymbol{x}).
\end{equation}
In terms of the transformed variables, we then have
\begin{equation}
    \Tilde{\boldsymbol{\nu}}(\boldsymbol{y})
    =
    \boldsymbol{J}(\boldsymbol{g}^{-1}(\boldsymbol{y}))
    \boldsymbol{\nu}(\boldsymbol{g}^{-1}(\boldsymbol{y}))
    \quad \mathrm{and} \quad
    \Tilde{p}(\boldsymbol{y})
    =
    \frac{1}{|\det(\boldsymbol{J}(\boldsymbol{g}^{-1}(\boldsymbol{y})))|}
    p (\boldsymbol{g}^{-1}(\boldsymbol{y})).
\end{equation}
We introduce the transformed version of the learning rate matrix:
\begin{equation}
    \Tilde{L}_{\alpha \beta}
    =
    \int d \boldsymbol{y}
    \Tilde{\nu}_{\alpha}(\boldsymbol{y})
    \partial_{y_{\beta}} \ln \Tilde{p}(\boldsymbol{y}) \Tilde{p}(\boldsymbol{y}).
\end{equation}
We have $l_i = L_{ii}$ and $\Tilde{l}_i=\Tilde{L}_{ii}$, which implies that this matrix is always trace-free and that this property is preserved under transformations. Following the above, the two matrices are related by
\begin{equation}
\begin{split}
    \Tilde{L}_{\alpha \beta}
    &=
    \int d \boldsymbol{x}
    \sum_{j,k}
    J_{\alpha j}(\boldsymbol{x})
    \nu_j (\boldsymbol{x})
    \partial_{x_k} \ln
    \left(
    \frac{p(\boldsymbol{x})}{|\det (\boldsymbol{J}(\boldsymbol{x}))|}
    \right) J^{-1}_{k\beta}(\boldsymbol{x})p(\boldsymbol{x})\\
    &=
    \int d \boldsymbol{x}
    \sum_{j,k}
    J_{\alpha j}(\boldsymbol{x})
    \nu_j (\boldsymbol{x})
    \partial_{x_k} \ln p(\boldsymbol{x}) J^{-1}_{k\beta}(\boldsymbol{x})p(\boldsymbol{x})\\
    &-
    \int d \boldsymbol{x}
    \sum_{j,k}
    J_{\alpha j}(\boldsymbol{x})
    \nu_j (\boldsymbol{x})
    \partial_{x_k} \ln |\det (\boldsymbol{J}(\boldsymbol{x}))|
    J^{-1}_{k\beta}(\boldsymbol{x})p(\boldsymbol{x}).
\end{split}
\end{equation}
Defining the matrices
\begin{equation}
    \mathcal{L}_{\alpha \beta}
    =
    \nu_{\alpha}(\boldsymbol{x})
    \partial_{x_\beta} \ln p(\boldsymbol{x})
    \quad \text{and} \quad
    \mathcal{M}_{\alpha \beta}(\boldsymbol{x})
    =
    \nu_{\alpha}(\boldsymbol{x}) \partial_{x_\beta} \ln |\det (\boldsymbol{J}(\boldsymbol{x}))|,
\end{equation}
we can write the original and transformed learning rate matrices as
\begin{equation}
    \boldsymbol{L}
    =
    \langle \mathcal{L} \rangle\quad \text{and}
    \quad
    \Tilde{L}
    =
    \langle \boldsymbol{J} \mathcal{L} \boldsymbol{J}^{-1} \rangle
    -
    \langle \boldsymbol{J} \mathcal{M} \boldsymbol{J}^{-1} \rangle.
\end{equation}
For the linear transformation
\begin{equation}
    \boldsymbol{y}
    =
    \boldsymbol{R} \boldsymbol{x},
\end{equation}
where $\boldsymbol{R}$ is a constant and invertible matrix, we obtain $\boldsymbol{J}=\boldsymbol{R}$ and $\boldsymbol{\mathcal{M}}=0$. Therefore, we obtain (\ref{transformation_learning_matrix}).

\section{Explicit expression of the learning rate matrix}
\label{appendix_linear_case}
Let us consider a special case of (\ref{Langevin_equation}) in which the system is linear. When the system is described by the linear Langevin equation, the force is assumed to be 
\begin{equation}
    \boldsymbol{F}_t(\boldsymbol{x})
    =
    -\boldsymbol{K}_t \boldsymbol{x} + \boldsymbol{k}_t,
\end{equation}
and the diffusion matrix $\boldsymbol{D}_t(\boldsymbol{x})=\boldsymbol{D}_t$ is independent of the coordinate. For a stable steady distribution, the drift matrix $\boldsymbol{K}_{\mathrm{st}}$ has eigenvalues with positive real parts. When the initial distribution is Gaussian, the probability density corresponding to (\ref{Fokker_Planck}) is Gaussian:
\begin{equation}
\label{two_steady_state_density}
    p_t(\boldsymbol{x})
    =
    \frac{1}{\sqrt{(2\pi)^N\det(\boldsymbol{\Xi}_t)}}
    \exp{
    \left(
    -\frac{1}{2}
    (\boldsymbol{x}-\boldsymbol{m}_t)^{\mathrm{T}} \boldsymbol{\Xi}_t^{-1} (\boldsymbol{x}-\boldsymbol{m}_t)
    \right)
    },
\end{equation}
where the mean $\boldsymbol{m}_t$ and covariance matrix $\boldsymbol{\Xi}_t$ are determined by the ordinary differential equations
\begin{equation}
    \dot{\boldsymbol{m}}_t
    =
    - 
    \boldsymbol{K}_t \boldsymbol{x}
    +
    \boldsymbol{k}_t,
\end{equation}
\begin{equation}
    \dot{\boldsymbol{\Xi}}_t
    =
    \boldsymbol{K}_t
    \boldsymbol{\Xi}_t
    +
    \boldsymbol{\Xi}_t
    \boldsymbol{K}_t^{\mathrm{T}}
    +
    2\boldsymbol{D}_t.
\end{equation}
The linear Langevin system maintains a Gaussian distribution at all times if and only if the initial distribution is Gaussian. This is because, when the initial distribution is Gaussian, all cumulants of order three or higher vanish at any time.
Using the expression for the time-derivative of the covariance matrix, we can write the local mean velocity as
\begin{equation}
\begin{split}
    \boldsymbol{\nu}_t(\boldsymbol{x})
    &=
    -\frac{1}{2} (\boldsymbol{K}_t
    \boldsymbol{\Xi}_t
    -
    \boldsymbol{\Xi}_t
    \boldsymbol{K}_t^{\mathrm{T}}
    -
    \dot{\boldsymbol{\Xi}}_t
    )
    \boldsymbol{\Xi}_t^{-1} 
    (\boldsymbol{x}-\boldsymbol{m}_t)
    +
    \dot{\boldsymbol{m}}\\
    &=
    -\frac{1}{2} (\boldsymbol{A}_t - \dot{\boldsymbol{\Xi}}_t)\boldsymbol{\Xi}_t^{-1} (\boldsymbol{x}-\boldsymbol{m}_t)
    +
    \dot{\boldsymbol{m}},
\end{split}
\end{equation}
where we have defined the skew-symmetric matrix
\begin{equation}
    \boldsymbol{A}_t
    =
    (\boldsymbol{K}_t
    \boldsymbol{\Xi}_t
    -
    \boldsymbol{\Xi}_t
    \boldsymbol{K}_t^{\mathrm{T}}).
\end{equation}
This is called irreversible circulation \cite{irreversible_circulation}, which quantifies the distance from equilibrium.\\
Then, we obtain the following explicit expression of the learning rate matrix:
\begin{equation}
\label{learning_rate_matrix_linear_system2}
    \boldsymbol{L}_t
    =
    \frac{1}{2}
    (\boldsymbol{A}_t-\dot{\boldsymbol{\Xi}}_t) \boldsymbol{\Xi}_t^{-1} + \dot{\boldsymbol{S}}_t.
\end{equation}
Now, the Fisher information is given by
\begin{equation}
\label{linear_Fisher}
    \mathcal{I}_t
    =
    \boldsymbol{\Xi}_t^{-1}.
\end{equation} 
Using (\ref{learning_rate_matrix_linear_system2}) and (\ref{linear_Fisher}), the entropy production rate (\ref{total_entropy_production_local}) is expressed as
\begin{equation}
\begin{split}
    \sigma_t
    &=
    \frac{1}{4}
    \mathrm{tr}
    \left(
    \boldsymbol{D}_t^{-1}
    (\boldsymbol{A}_t-\dot{\boldsymbol{\Xi}}_t)
    \boldsymbol{\Xi}_t^{-1}
    (\boldsymbol{A}_t-\dot{\boldsymbol{\Xi}}_t)^{\mathrm{T}}
    \right)
    +
    \dot{\boldsymbol{m}}_t^{\mathrm{T}}
    \boldsymbol{D}_t^{-1}
    \dot{\boldsymbol{m}}_t\\
    &=
    \mathrm{tr}
    \left(
    \boldsymbol{D}^{-1}_t
    \left(\boldsymbol{L}_t - \dot{\boldsymbol{S}}_t\right)
    \boldsymbol{\Xi}_t
    \left(\boldsymbol{L}_t - \dot{\boldsymbol{S}}_t\right)^{\mathrm{T}}
    \right)
    +
    \dot{\boldsymbol{m}}_t^{\mathrm{T}}
    \boldsymbol{D}_t^{-1}
    \dot{\boldsymbol{m}}_t\\
    &=
    \mathrm{tr}
    \left(
    \boldsymbol{D}^{-1}_t
    \left(\boldsymbol{L}_t - \dot{\boldsymbol{S}}_t\right)
    \boldsymbol{\mathcal{I}}_t^{-1}
    \left(\boldsymbol{L}_t - \dot{\boldsymbol{S}}_t\right)^{\mathrm{T}}
    \right)
    +
    \dot{\boldsymbol{m}}_t^{\mathrm{T}}
    \boldsymbol{D}_t^{-1}
    \dot{\boldsymbol{m}}_t.
\end{split}
\end{equation}
Thus, for linear dynamics, the lower bound (\ref{total_entropy_production_lower_bound}) becomes an equality. In the steady state, where $\dot{\boldsymbol{m}}_t,\dot{\boldsymbol{S}}_t,\dot{\boldsymbol{\Xi}}_t$ all vanish, we have the identity
\begin{equation}
    \sigma_{\mathrm{st}}
    =
    \mathrm{tr}
    \left(
    \boldsymbol{D}_{\mathrm{st}}^{-1}
    \boldsymbol{L}_{\mathrm{st}}
    \mathcal{I}_{\mathrm{st}}^{-1}
    \boldsymbol{L}_{\mathrm{st}}^{\mathrm{T}}
    \right).
\end{equation}

\section{Derivation of (\ref{learning_matrix_Fisher_local_velocity})}
\label{appendix_learning_matrix_Fisher_local_velocity}
We can use the Cauchy-Schwarz inequality to relate the learning rate matrix to the entropy production rate. First, we use
\begin{equation}
    \begin{split}
        \left(
        \boldsymbol{a}^\mathrm{T}
        \left(
        \boldsymbol{L}_t
        -
        \dot{\boldsymbol{S}}_t
        \right)
        \boldsymbol{b}
        \right)^2
        &=
        \left(
        \sum_{k,l}
        a_k
        \left(
        (\boldsymbol{L}_t)_{kl}
        -
        \dot{S}_t^k\delta_{kl}
        \right)
        b_l
        \right)^2\\
        &=
        \left(
        \langle
        \sum_k a_k (\nu_{t,k}-\langle \nu_{t,k}\rangle)
        \sum_l b_l \partial_{x_l} \ln p_t
        \rangle
        \right)^2\\
        &\leq
        \sum_{k,l}
        \langle
        a_k (\nu_{t,k}-\langle \nu_{t,k}\rangle) a_l (\nu_{t,l}-\langle \nu_{t,l}\rangle)\rangle
        \sum_{m,n}
        \langle
        b_m \partial_{x_m} \ln p_t
        b_n \partial_{x_n} \ln p_t
        \rangle\\
        &=
        \boldsymbol{a}^{\mathrm{T}}
        \boldsymbol{\Xi}_{\nu,t}
        \boldsymbol{a}
        \boldsymbol{b}^{\mathrm{T}}
        \mathcal{I}_t
        \boldsymbol{b}.
    \end{split}
\end{equation}
We can write the above inequality as
\begin{equation}
    \boldsymbol{a}^{\mathrm{T}}
    \boldsymbol{\Xi}_{\nu,t}
    \boldsymbol{a}
    \geq
    \frac{\left(\boldsymbol{a}^\mathrm{T}
    \left(
    \boldsymbol{L}_t
    -
    \dot{\boldsymbol{S}}_t
    \right)
    \boldsymbol{b}\right)^2}{\boldsymbol{b}^{\mathrm{T}}
    \mathcal{I}_t
    \boldsymbol{b}}
    =
    \frac{\boldsymbol{a}^\mathrm{T}
    \left(
    \boldsymbol{L}_t
    -
    \dot{\boldsymbol{S}}_t
    \right)
    \boldsymbol{b}
    \boldsymbol{b}^\mathrm{T}
    \left(
    \boldsymbol{L}_t
    -
    \dot{\boldsymbol{S}}_t
    \right)^{\mathrm{T}}
    \boldsymbol{a}
    }{\boldsymbol{b}^{\mathrm{T}}
    \mathcal{I}_t
    \boldsymbol{b}}.
\end{equation}
Setting $\boldsymbol{b}=\mathcal{I}_t^{-1}\left(\boldsymbol{L}_t-\dot{\boldsymbol{S}}_t\right)^{\mathrm{T}}\boldsymbol{a}$, we obtain 
\begin{equation}
\begin{split}
    \boldsymbol{a}^{\mathrm{T}}
    \boldsymbol{\Xi}_{\nu,t}
    \boldsymbol{a}
    &\geq
    \frac{
    \left(
    \boldsymbol{a}^\mathrm{T}
    \left(
    \boldsymbol{L}_t
    -
    \dot{\boldsymbol{S}}_t
    \right)
    \boldsymbol{\mathcal{I}}_t^{-1}
    \left(
    \boldsymbol{L}_t
    -
    \dot{\boldsymbol{S}}_t
    \right)^{\mathrm{T}}
    \boldsymbol{a}
    \right)^2
    }{\left(
    \boldsymbol{a}^\mathrm{T}
    \left(
    \boldsymbol{L}_t
    -
    \dot{\boldsymbol{S}}_t
    \right)
    \boldsymbol{\mathcal{I}}_t^{-1}
    \left(
    \boldsymbol{L}_t
    -
    \dot{\boldsymbol{S}}_t
    \right)^{\mathrm{T}}
    \boldsymbol{a}
    \right)}\\
    &=
    \left(
    \boldsymbol{a}^\mathrm{T}
    \left(
    \boldsymbol{L}_t
    -
    \dot{\boldsymbol{S}}_t
    \right)
    \boldsymbol{\mathcal{I}}_t^{-1}
    \left(
    \boldsymbol{L}_t
    -
    \dot{\boldsymbol{S}}_t
    \right)^{\mathrm{T}}
    \boldsymbol{a}
    \right).
\end{split}
\end{equation}
Because this is valid for any vector $\boldsymbol{a}$, we have matrix inequality (\ref{learning_matrix_Fisher_local_velocity}). We can also derive (\ref{learning_matrix_Fisher_local_velocity}) using the Schur complement: The matrix 
\begin{equation}
    \left(
    \begin{matrix}
        \boldsymbol{\Xi}_t& (\boldsymbol{L}_t-\dot{\boldsymbol{S}}_t)\\
        (\boldsymbol{L}_t-\dot{\boldsymbol{S}}_t)^{\mathrm{T}}&\boldsymbol{\mathcal{I}}_t
    \end{matrix}
    \right)
\end{equation}
is positive definite, which directly yields (\ref{learning_matrix_Fisher_local_velocity}).

\section{Derivations of lower and tighter bounds}
\label{appendix_tighter_bound}
A tighter bound can be obtained using the relation
\begin{equation}
\begin{split}
    &l^y_{\mathrm{st}}
    =
    \left\langle
    \boldsymbol{\nu}^{y,T}_{\mathrm{st}}
    \boldsymbol{\nabla}_y
    \ln p_{\mathrm{st}}
    \right\rangle
    =
    \left\langle
    \boldsymbol{\nu}^{y,T}_{\mathrm{st}}
    \boldsymbol{\nabla}_y
    \ln p_{\mathrm{st}}^{|y}
    \right\rangle\\
    &\Rightarrow
    (l^y_{\mathrm{st}})^2
    \leq
    \left(
    \sigma^y_{\mathrm{st}}
    -
    \langle\dot{y}\rangle^T_{\mathrm{st}} (\boldsymbol{\mu}_{\mathrm{st}}^yT^y)^{-1}
    \langle\dot{y}\rangle^T_{\mathrm{st}}
    \right)
    \mathrm{tr} (\boldsymbol{\mu}_{\mathrm{st}}^yT^y \boldsymbol{\mathcal{I}}^{|y}_{\mathrm{st}}),
\end{split}
\end{equation}
where $\boldsymbol{\mathcal{I}}^{|y}_{\mathrm{st}}$ is the Fisher information matrix corresponding to the conditional probability density $\ln p_{\mathrm{st}}^{|y}(\boldsymbol{z}|\boldsymbol{y})$, and we set $\boldsymbol{D}_{\mathrm{st}}^y=\boldsymbol{\mu}_{\mathrm{st}}^y T^y$. Recalling the chain rule for the Fisher information expressed as
\begin{equation}
    \boldsymbol{\mathcal{I}}^{yy}_{\mathrm{st}}
    =
    \boldsymbol{\mathcal{I}}^{y}_{\mathrm{st}}
    +
    \boldsymbol{\mathcal{I}}^{|y}_{\mathrm{st}}
    \quad \text{with}
    \quad
    (\boldsymbol{\mathcal{I}}^{y}_{\mathrm{st}})_{kl}
    =
    \langle
    \partial_{y_k}
    \ln p^y_{\mathrm{st}}
    \partial_{y_l}
    \ln p^y_{\mathrm{st}}
    \rangle,
\end{equation}
we find that $\boldsymbol{\mathcal{I}}^{yy}_{\mathrm{st}} \geq \boldsymbol{\mathcal{I}}^{|y}_{\mathrm{st}}$, and we have the tighter bound
\begin{equation}
    \dot{Q}^y_{\mathrm{st}}
    \geq
    T^y l^y_{\mathrm{st}}
    +
    \frac{(l^y_{\mathrm{st}})^2}{\mathrm{tr}\left(
    \boldsymbol{\mu}_{\mathrm{st}}^y \boldsymbol{\mathcal{I}}^{|y}_{\mathrm{st}}
    \right)}
    +
    \langle
    \dot{y}
    \rangle^\mathrm{T}_{\mathrm{st}}
    (\boldsymbol{\mu}_{\mathrm{st}}^y)^{-1}
    \langle
    \dot{y}
    \rangle^\mathrm{T}_{\mathrm{st}},
\end{equation}
where the heat dissipation from the subsystem $\boldsymbol{y}$ to the environment in the steady state is defined as $\dot{Q}^y_{\mathrm{st}}=\dot{S}^{y,\mathrm{env}}_{\mathrm{st}}/T^y$.
Minimizing the right-hand side with respect to $l^y_{\mathrm{st}}$ yields a global lower bound on the dissipation of a subsystem:
\begin{equation}
    \dot{Q}^y_{\mathrm{st}}
    \geq
    -\frac{(T^y)^2}{4}
    \mathrm{tr}
    \left(
    \boldsymbol{\mu}_{\mathrm{st}}^y
    \boldsymbol{\mathcal{I}}^{|y}_{\mathrm{st}}
    \right)
    +
    \langle
    \dot{y}
    \rangle^\mathrm{T}_{\mathrm{st}}
    (\boldsymbol{\mu}_{\mathrm{st}}^y)^{-1}
    \langle
    \dot{y}
    \rangle^\mathrm{T}_{\mathrm{st}}
    \geq
    -\frac{(T^y)^2}{4}
    \mathrm{tr}
    \left(
    \boldsymbol{\mu}_{\mathrm{st}}^y
    \boldsymbol{\mathcal{I}}^{|y}_{\mathrm{st}}
    \right).
\end{equation}
This implies that the conditional Fisher information, which measures the degree of correlation between $\boldsymbol{y}$ and $\boldsymbol{z}$, provides a general limit on the magnitude of the apparent violation of the second law in a subsystem. The dissipation rate depends on the flows in the system, whereas the bound depends only on the steady state correlations. Thus, the correlations encoded in the steady state probability density determine the maximum apparent violation of the second law, independent of how far the system is from equilibrium. In the time-dependent case, we find the bounds
\begin{equation}
    \dot{S}^{y,\mathrm{env}}_t
    +
    \dot{S}^{y}_t
    -
    l^y_{t}
    \geq
    \frac{
    \left(\dot{S}^{y}_t-
    l^y_{t}
    \right)^2
    }
    {
    \mathrm{tr}
    \left(
    \boldsymbol{\mu}^y_t  \boldsymbol{\mathcal{I}}_t^{yy}
    \right)}
    +
    \langle\boldsymbol{\dot{y}}^y\rangle^{\mathrm{T}}
    (\boldsymbol{\mu}^y_t)^{-1}
    \langle\boldsymbol{\dot{y}}^y\rangle,
\end{equation}
\begin{equation}
    \dot{S}^{y,\mathrm{env}}_t
    +
    \dot{S}^{y}_t
    -
    l^y_{t}
    \geq
    \frac{
    \left(
    l^y_{t}
    \right)^2
    }
    {
    \mathrm{tr}
    \left(
    \boldsymbol{\mu}^y_t  \boldsymbol{\mathcal{I}}_t^{|y}
    \right)}
    +
    \langle\boldsymbol{\dot{y}}^y\rangle^{\mathrm{T}}
    (\boldsymbol{\mu}^y_t)^{-1}
    \langle\boldsymbol{\dot{y}}^y\rangle.
\end{equation}
Minimizing with respect to the learning rate yields the global lower bounds
\begin{equation}
    \dot{Q}^y_t
    \geq
    -\frac{(T^y_t)^2}{4} 
    \mathrm{tr}
    \left(
    \boldsymbol{\mu}^y_t
    (\boldsymbol{\mathcal{I}}^y_t
    +
    \boldsymbol{\mathcal{I}}^{|y}_t)
    \right)
    \quad
    \text{and}
    \quad
    \dot{Q}^y_t + T^y_t\dot{S}^y_t
    \geq
    -\frac{(T^y_t)^2}{4} 
    \mathrm{tr}
    \left(
    \boldsymbol{\mu}^y_t
    \boldsymbol{\mathcal{I}}^{|y}_t
    \right).
\end{equation}
The second inequality reproduces the second law in the absence of correlations between $\boldsymbol{y}$ and $\boldsymbol{z}$ and   in the steady state. However, the first inequality gives a new lower bound on the dissipation, even in the absence of correlations (or for the total system):
\begin{equation}
    \dot{Q}_t
    \geq
    -T_t \dot{S}_t
    +
    \frac{(\dot{S}_t)^2}{\mathrm{tr}\left(\boldsymbol{\mu}_t \boldsymbol{\mathcal{I}_t}\right)}
    \geq
    -\frac{(T_t)^2}{4} 
    \mathrm{tr}
    \left(
    \boldsymbol{\mu}_t
    \boldsymbol{\mathcal{I}}_t
    \right),
\end{equation}
where the heat dissipation from the total system to the environment is defined as $\dot{Q}_t=\dot{S}^{\mathrm{env}}_t/T_t$.
Similar to the relation involving the learning rate, this implies that, while an increase in Shannon entropy can allow a time-dependent system to have a negative rate of dissipation, a very rapid increase in Shannon entropy will always lead to a positive dissipation rate. The maximal negative dissipation rate is determined by the Fisher information matrix, that is, the spatial structure of the probability density. For Gaussian systems, we have $\boldsymbol{\mathcal{I}}_t=\boldsymbol{\Xi}_t^{-1}$, and so
\begin{equation}
    \dot{Q}_t
    \geq
    -T_t \dot{S}_t
    +
    \frac{(\dot{S}_t)^2}{\mathrm{tr}\left(\boldsymbol{\mu}_t \boldsymbol{\Xi}_t^{-1}\right)}
    \geq
    -\frac{(T_t)^2}{4} 
    \mathrm{tr}
    \left(
    \boldsymbol{\mu}_t
    \boldsymbol{\Xi}_t^{-1}
    \right).
\end{equation}
Therefore, achieving a large negative dissipation rate necessarily requires a narrow probability density realized by tight control of the dynamics.

\section{Smallest possible diagonal
element}
\label{smallest_eigenvalue}
Let $\boldsymbol{d}$ denote the vector of diagonal elements of the symmetric part of the learning rate matrix $\tilde{\boldsymbol{L}}^{\mathrm{sym}}$, arranged in non-increasing order $(d_{\mathrm{max}}=d_1\geq d_2 \geq \cdots \geq d_N=d_{\mathrm{min}})$, in a new coordinate system obtained by an orthogonal transformation.
Let $\boldsymbol{\lambda}$ denote the eigenvalues of the original symmetric learning rate matrix $\boldsymbol{L}^{\mathrm{sym}}$, also arranged in non-increasing order $(\lambda_{\mathrm{max}}=\lambda_1\geq \lambda_2 \geq \cdots \geq \lambda_N=\lambda_{\mathrm{min}})$.
According to the Schur–Horn theorem,
\begin{equation}
\begin{split}
    &\sum_{i=1}^{n}
    d_i 
    \leqq
    \sum_{i=1}^{n}
    \lambda_i\quad \mathrm{for}\quad n=1,\cdot\cdot\cdot,N-1,\\
    &\sum_{i=1}^{N}
    d_i 
    =
    \sum_{i=1}^{N}
    \lambda_i,
\end{split}
\end{equation}
we have
\begin{equation}
\begin{split}
    &\sum_{i=1}^{N}d_i 
    =
    \sum_{i=1}^{N}
    \lambda_i,\\
&\sum_{i=1}^{N-1}d_i +d_{\mathrm{min}}
    =
    \sum_{i=1}^{N-1}
    \lambda_i +\lambda_{\mathrm{min}} \geq  \sum_{i=1}^{N-1}
    d_i +\lambda_{\mathrm{min}},\\
&\Rightarrow \lambda_{\mathrm{min}} \leq d_{\mathrm{min}},
\end{split}
\end{equation}
where we use the Shur-Horn theorem at the second line. The equality is achieved when the orthogonal matrix diagonalizes $\boldsymbol{L}^{\mathrm{sym}}$. Therefore, we obtain
\begin{equation}
\min_{R \in O(N)} \left( \min_i \left( R \boldsymbol{L} R^T\right)_{ii} \right)  = \lambda_{\min}.
\end{equation}
Therefore, the eigenvalues give the smallest possible diagonal element.


\section*{References}
    \begin{thebibliography}{99}
        \bibitem{Sagawa} Sagawa T and Ueda M 2012 \textit{Phys. Rev. E} \textbf{85} 021104
        \bibitem{parrondo2015thermodynamics} Parrondo J M R, Horowitz J M and Sagawa T 2015 \textit{Nature Physics} \textbf{11} 131
        \bibitem{coverthomas} Cover T M 1999 \textit{Elements of Information Theory} (New York: Wiley)
        \bibitem{Sekimoto2010} Sekimoto K 2010 Concept of Heat on Mesoscopic Scales in \textit{Stochastic Energetics} (Berlin: Springer) pp. 135--174
        \bibitem{cellar_information} Barato A C, Hartich D and Seifert U 2014 \textit{New J. Phys.} \textbf{16} 103024
        \bibitem{Sensory_capacity} Hartich D, Barato A C and Seifert U 2016 \textit{Phys. Rev. E} \textbf{93} 022116
        \bibitem{ito2015maxwell} Ito S and Sagawa T 2015 \textit{Nature Communications} \textbf{6} 7498
        \bibitem{TUR} Barato A C and Seifert U 2015 \textit{Phys. Rev. Lett.} \textbf{114} 158101
        \bibitem{Tanogami_Saito_2023} Tanogami T, Van Vu T and Saito K 2023 \textit{Phys. Rev. Res.} \textbf{5} 043280
        \bibitem{langevin_information} Allahverdyan A E, Janzing D and Mahler G 2009 \textit{J. Stat. Mech.} \textbf{2009} P09011
        \bibitem{conti_information} Horowitz J M and Esposito M 2014 \textit{Phys. Rev. X} \textbf{4} 031015
        \bibitem{Multipartite_information_flow} Horowitz J M 2015 \textit{J. Stat. Mech.} \textbf{2015} P03006
        \bibitem{penocchio2022information} Penocchio E, Avanzini F and Esposito M 2022 \textit{J. Chem. Phys.} \textbf{157} 3
        \bibitem{non-bipartite} Chétrite R, Rosinberg M L, Sagawa T and Tarjus G 2019 \textit{J. Stat. Mech.} \textbf{2019} 114002
        \bibitem{gardiner1985handbook} Gardiner C W 1985 \textit{Handbook of Stochastic Methods} (Berlin: Springer)
        \bibitem{Shannon} Shannon C E 1948 \textit{Bell Syst. Tech. J.} \textbf{27} 379
        \bibitem{SUNHAN198026} Han T S 1980 \textit{Information and Control} \textbf{46} 26
        \bibitem{irreversible_circulation} Tomita K and Tomita H 1974 \textit{Prog. Theor. Phys.} \textbf{51} 1731
        \bibitem{Horn_Johnson_1985} Horn R A and Johnson C R 1985 \textit{Matrix Analysis} (Cambridge: Cambridge University Press)
        \bibitem{Brownian_gyrator} Filliger R and Reimann P 2007 \textit{Phys. Rev. Lett.} \textbf{99} 230602
        \bibitem{Information_Arbitrage} Leighton M P, Ehrich J and Sivak D A 2024 arXiv:2308.06325
        \bibitem{two_interaction} Rieser J, Ciampini M A, Rudolph H, Kiesel N, Hornberger K, Stickler B A, Aspelmeyer M and Delić U 2022 \textit{Science} \textbf{377} 987
    \end{thebibliography}
\end{document}